\documentclass{article}
\pdfoutput=1
\usepackage{amsmath,amssymb,graphicx,microtype,hyperref}

\def\be{\begin{eqnarray}}
\def\ee{\end{eqnarray}}
\def\nn{\nonumber}

\def\p{\partial}

\def\ttau{p}
\def\intertwin{{\cal G}}

\def\l[{\phantom.[}

\hoffset= - 1.0in         

\begin{document}

\title{\vspace{.1cm}{\Large {\bf Ding-Iohara-Miki symmetry of network matrix models}\vspace{.2cm}}
\author{{\bf A. Mironov$^{a,b,c,d,}$}\footnote{mironov@lpi.ru; mironov@itep.ru}, \ {\bf A. Morozov$^{b,c,d,}$}\thanks{morozov@itep.ru}, \ and \ {\bf Y. Zenkevich$^{b,d,e,}$}\thanks{yegor.zenkevich@gmail.com}}
\date{ }
}

\maketitle

\vspace{-5.5cm}

\begin{center}
\hfill FIAN/TD-07/16\\
\hfill IITP/TH-05/16\\
\hfill ITEP/TH-06/16\\
\hfill INR-TH-2016-008
\end{center}

\vspace{3.2cm}

\begin{center}
$^a$ {\small {\it Lebedev Physics Institute, Moscow 119991, Russia}}\\
$^b$ {\small {\it ITEP, Moscow 117218, Russia}}\\
$^c$ {\small {\it Institute for Information Transmission Problems, Moscow 127994, Russia}}\\
$^d$ {\small {\it National Research Nuclear University MEPhI, Moscow 115409, Russia }}\\
$^e$ {\small {\it Institute of Nuclear Research, Moscow 117312, Russia }}
\end{center}

\vspace{.5cm}

\begin{abstract}
  Ward identities in the most general ``network matrix model'' from
  \cite{ellMMZ} can be described in terms of the Ding-Iohara-Miki
  algebras (DIM).  This confirms an expectation that such algebras and
  their various limits/reductions are the relevant
  substitutes/deformations of the Virasoro/W- algebra for $(q,t)$ and
  $(q_1,q_2,q_3)$ deformed network matrix models.  Exhaustive for
  these purposes should be the {\it Pagoda} triple-affine elliptic
  DIM, which corresponds to networks associated with 6d gauge theories
  with adjoint matter (double elliptic systems). We provide some
  details on elliptic $qq$-characters.
\end{abstract}

\vspace{.5cm}

\section{Introduction}

Recently,
basing on the previous studies in
\cite{UFN23}-\cite{MMZ},
we introduced \cite{ellMMZ} a generic Dotsenko-Fateev (DF)  \cite{DF}
network conformal matrix model, associated with the most general brane web/network
(the low-energy limit of toric Calabi-Yau compactifications).
The first question to ask about this theory
is what is the set of the relevant ``Virasoro/W- constraints'':
the Ward identities, which are satisfied by its partition function.
In this paper, we argue that the substitute/deformation of
the CFT stress tensor, which generates these identities,
is now provided by the analogues of the $q$-characters
\cite{FR} in the elliptic Ding-Iohara-Miki algebra (DIM) \cite{DIM,FFJMM},
as anticipated at different deformation levels in \cite{Ja}-\cite{Mat}.

\bigskip

We remind \cite{UFN23} that there are three equivalent ways
to derive Ward identities in matrix models (and other quantum field and string theory models):

(i) by making a change of integration variable \cite{MMVir},

(ii) by considering an average of a total derivative \cite{loopeq} and

(iii) by building a matrix model from a free field correlator
with a given symmetry \cite{confmamo,MMMM}.

\noindent
The first two methods can seem identical, but in fact this is not quite true:
(ii) is technically simpler (more straightforward) than (i),
but instead the emerging algebraic structure is more difficult to reveal.
Ideal for this task is the method (iii), which we now briefly remind.

\bigskip

Given a symmetry generating operator (or a set of operators) $\hat T$
(say, the stress tensor and higher $W$-algebra generators),
one gets a set of identities
\be
\left< \Psi\Big|\ \hat \intertwin(\ttau) \ \hat T \ \hat Q \ \Big| {\rm vac} \right>\ = 0
\ \ \ \Longleftrightarrow \ \ \
L(\p_\ttau) \left< \Psi\Big|\ \hat \intertwin(\ttau) \ \hat Q \ \Big| {\rm vac} \right>\ = 0
\ee
where

$\bullet$ $|{\rm vac}\rangle\ $ is a ``vacuum'' state annihilated by $\hat T$,
$\ \hat T\,|{\rm vac} \rangle\ = 0$,

$\bullet$ $\hat Q$ is any ``screening'' operator
which commutes with $\hat T$, $\ [\hat T,\hat Q] = 0$,

$\bullet$ $\langle\Psi|$ is an arbitrary state usually made out of
vertex operators, and

$\bullet$ $\hat \intertwin(\ttau)$ is an intertwiner with the property
$L(\p_\ttau)\hat \intertwin(\ttau) = \hat G(\ttau)\hat T$, which
can be used to convert operator(s) $\hat T$ into differential/difference
operators $L$ acting on ``the time variables'' $\ttau$.

This is a very general group theoretical construction
describing a partition function
\be
Z(\ttau)\  =\ \left< \Psi\Big|\ \hat \intertwin(\ttau) \ \hat Q \ \Big| {\rm vac} \right>
\ee
with a given ($T$-induced) symmetry as a matrix element.
Conformal matrix models \cite{confmamo} arise in this way when matrix elements
are correlators of $2d$ free fields
and integrals or sums over partitions
(interpreted as matrix model eigenvalue integrals/sums)
emerge from an explicit description of screening charges $\hat Q$
(which are the centralizer of $\hat T$) in the free field Fock space.

\bigskip

\begin{figure}[h!]
  \centering
    \includegraphics[width=12cm]{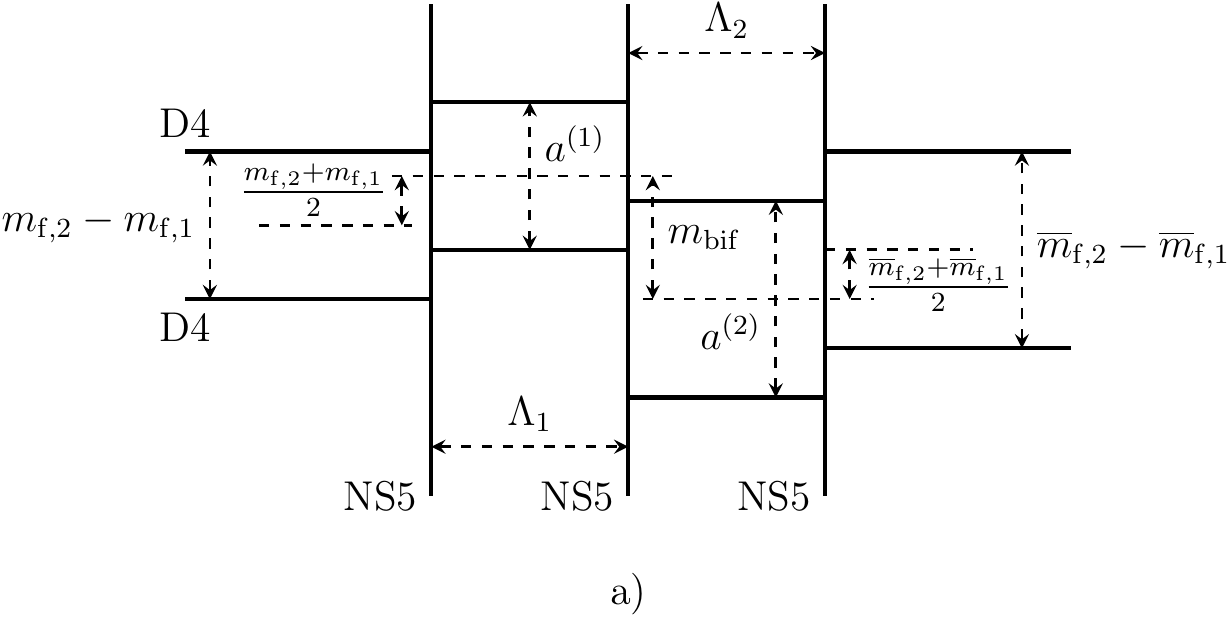}\\
  \includegraphics[width=9cm]{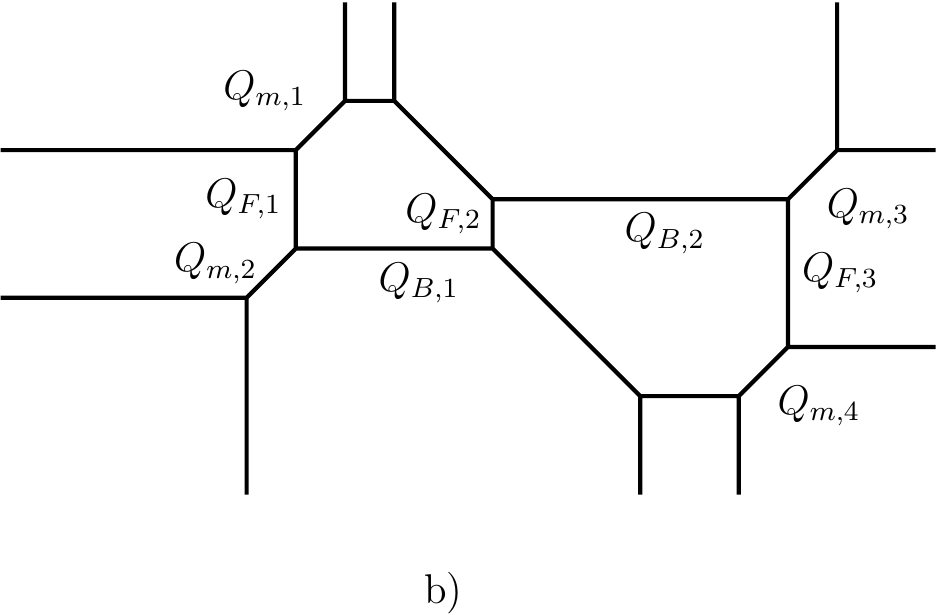} \qquad \includegraphics[width=6cm]{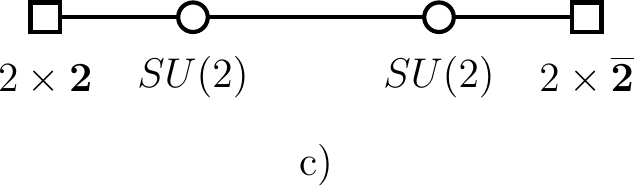}
  \caption{\footnotesize{a) Type IIA brane diagram consisting of
    two horizontal and three vertical intersecting lines representing
    NS5 and D4 branes. The low energy theory in this background is
    4d $\mathcal{N}=2$ gauge theory with $SU(2)^2$ gauge
    group. $\Lambda_i$ are exponentiated complexified gauge couplings,
    $a^{(a)}$ are Coulomb moduli and $m_a$ are the hypermultiplet
    masses. b) The toric diagram of the Calabi-Yau threefold,
    corresponding to the $5d$ gauge theory with the same matter
    content. Edges represent two-cycles with complexified K\"ahler
    parameters $Q_i$, which play the same role as the distances
    between the branes in a). c) The quiver encoding the matter
    content of the gauge theory. $SU(2)$ gauge groups live on each
    node and bifundamental matter on each edge. The squares represent
    pairs of (anti)fundamental matter hypermultiplets.}}
  \label{fig:1}
\end{figure}

Reversing the logic, one can start from generic network matrix model
\cite{ellMMZ}, associated with the toric diagram in Fig.\ref{fig:1}
b),
\begin{equation}
Z_\Gamma(q,t|\ttau) = \sum_{\{R_E\}}  \prod_V
C_{R(E'_V),R(E''_V),R(E'''_V)}(q,t|\ttau_V)
\label{top-network}
\end{equation}
associated with a planar 3-valent graph $\Gamma$ with triples of edges
$E'_V,E''_V,E'''_V$ merging at vertices $V$.  The sum goes over Young
diagrams $R_E$ on the edges and the topological vertices $C$ are
provided \cite{ref,IKV,topvert} by weighted sums over $3d$
partitions with given boundary conditions
$R(E'_V),R(E''_V),R(E'''_V)$. The weights depend on compactification
parameters $q_{1,2,3}$ and on auxiliary time
variables $\ttau_V$ (their background values can be used to develop the check-operator formalism a la \cite{AMM}). Usually, time variables are ascribed to edges, not vertices (and we do so in (\ref{netmod}) below, but the right procedure remains disputable -- and in (\ref{top-network}) we absorb propagators into vertices to simplify the formula.  Two of the deformation parameters are also parameterized
as $q=e^{\epsilon_1},t=e^{-\epsilon_2}$ and the third one is put equal
to $q_3=t/q$ in 5d theory, while in 6d theories it remains free. In
fact, at least, at the algebraical level, more Kerov parameters \cite{Kerov} of the
same nature can also be included~\cite{IKV}. Especially
important is the elliptic case, where infinite set of Kerov parameters
form a geometric progression. For generic deformation the model
belongs to the universality class \cite{SW,GKMMM} of the double elliptic
integrable systems \cite{dell} and is invariant under a rich set of
large canonical transformations (dualities).  As to the infinitesimal
transformations, they also form an interesting closed algebra of Ward
identities, which can be revealed by the following sequence of steps
briefly encountered above:

\begin{itemize}
\item {\bf Breaking horizontal/vertical symmetry.}  Rewrite $Z_\Gamma$
  as an eigenvalue matrix integral/sum of the form ($\prod^{\prime}$
  means that the multiplier $\Delta\Big(x_{i,\alpha}^{(a)},\vec
  x_{i,\alpha}^{(a)}\Big)$ is excluded from the product) \be
  Z_{\Gamma_{_{\vdash}}}(q|\ttau) = \int \left[\prod_{\alpha,a} d\vec
    x_{\alpha}^{(a)}\right]
  {\prod}^{\prime}_{{\alpha,\beta}\atop{a,b}}\Delta\Big(\vec
  x_{\alpha}^{(a)},\vec x_{\beta}^{(b)}\Big)^{C_{\alpha\beta}^{ab}}\
  \exp\left(\sum_{\alpha,a,k} \ttau_{k,\alpha}^a u_k(\vec
    x_\alpha^{(a)})\right)
\label{netmod}
\ee Here (Jackson like) integrals over $\vec x_\alpha =
\{x_{\alpha,i}^{(a)},\ i=1,\ldots,N_\alpha^{(a)},\
\alpha=1,\ldots,m^{(a)},\ a=1,\ldots,n\}$ substitute the sums over
Young diagrams for ``vertical'' edges $E^{(a)}_{\mathrm{vert}}$ of the
web-diagram $\Gamma$ (the choice is actually dictated by the $5d/4d$
limit and $N_\alpha^{(a)}$ may be interpreted as the number of lines
in the Young diagram, which can be arbitrary). In the example from
Fig.~\ref{fig:1} we have $n=2$, $m_1 = m_2 = 2$. We put an additional
index on $\Gamma_{\!_\vdash}$ to remind about additional
vertical/horizontal (quiver) structure on the graph $\Gamma$, implicit
in the formula. The set of ``Casimir'' functions $u_k(\vec x)$ is
usually adjusted to simplify the differential/difference equations
(\ref{Leq}) below (we will actually use the Miwa transform, converting
Casimirs into vertex operator insertions).  The sums over diagrams for
``horizontal'' edges $E_{\mathrm{hor}}^{(ab)}$ are substituted by the
$q_{1,2,3}$-dependent Vandermonde-like quantities $\Delta(\vec
x_\alpha^{(a)},\vec x_\beta^{(b)})$ which can be realized as a free
field pairwise correlator of {\bf screening currents}
$S_\alpha^{(a)}(x)$, and the product arises as a consequence of the
Wick theorem.

\item {\bf Screening operators.}
Using the screening charges given by
single-variable integrals (perhaps, Jackson sums)
of the screening currents (which are exponentials of free fields)
\be
\hat S_\alpha^{(a)} =\oint_{C_a} dx:\exp\left(\sum_{k\in \mathbb{Z}}x^k \hat a_{k,\alpha}\right):
\ \ \ \ [\hat a_{n,\alpha},\hat a_{m,\beta}]=\xi_n\delta_{n+m,0}\delta_{\alpha\beta}
\ee
one can realize the matrix model integral as an average
\begin{equation}
Z_{\Gamma_{\!_\vdash} }(\ttau)=\ \left\langle 0 \Bigg| \,\hat \intertwin(\ttau)\
\overbrace{\prod_{\alpha,a} \Big(\hat S_\alpha^{(a)}\Big)^{N_{\alpha}^{(a)}}}^{\hat Q}\,
\Bigg| 0\right\rangle_N
\end{equation}
where the vacuum state is $N=\sum_\alpha N_\alpha^{(a)}$-charged
vacuum with respect to the Heisenberg operators $\hat a_{n,\alpha}$
and $\hat G(t)$. The number of free fields $\{a_{n,\alpha}\}$ actually
depends on the number of horizontal edges (``D-branes'') in the
original graph $\Gamma$, i.e.\ on the ranks of gauge groups in $4d$
version of the model (\ref{netmod}).  We implicitly include $N$ into
the set of time-variables.

\item {\bf Ward identities} satisfied by the matrix integral can be
  described in two ways.  In the free field terms they are provided by
  the free field operators $\hat {W}_a$ that are defined to commute
  with the screening charges, while in terms of time variables (i.e.\
  literally as a set of constraints imposed on the time-dependent
  integral) they are expressed with the help of the intertwiner $\hat
  \intertwin(\ttau)$:
  \be
  {W}_a(\ttau,\partial_\ttau)Z_{\Gamma_{\!_\vdash}}(\ttau)\ = \
  \left<0\Big|\ \hat \intertwin(\ttau)\ \hat { W}_a \ \hat Q \
    \Big|0\right>_N \ = \ 0
\label{Leq}
\ee
The question is what is the algebra formed by this set of
constraints on a matrix integral.  In simplest examples this is just a
Borel subalgebra of Virasoro or various $W_{m_{a}}$ algebras, where
${m_a}$ are related by the number of horizontal edges in $\Gamma$
(e.g.\ $m_1 = m_2 = 2$ in Fig.~\ref{fig:1}).

\item {\bf Toroidal algebra.}  One can embed all the
  $W_{m_a}$-algebras associated with the set of matrix integrals of a
  given type into a larger algebra. For instance, in the case of
  Dotsenko-Fateev integrals associated with Nekrasov functions and
  topological vertices corresponding to all graphs, this gives rise to
  toroidal algebras: affine Yangians in the case of 4d Nekrasov
  functions \cite{4dDIM,Mat}, Ding-Iohara-Miki (DIM) (quantum
  toroidal) algebra in the 5d case \cite{DIM,FFJMM,F} and elliptic DIM
  algebra in the 6d case \cite{FHHSY,Iq}. Concrete quiver corresponds to
  a set of representations of the toroidal algebra given by a fixed
  number of Young diagrams. Moreover, the (refined) topological vertex
  can be obtained as a matrix element of the intertwining operators of
  the DIM algebra \cite{AF}. We concentrate below on the level 1
  representation of DIM algebra so that there always exists a simple
  bosonization \cite{FHHSY}. For generic levels an analogue of
  the free-field representation of Kac-Moody algebras \cite{WZNW} will
  be needed.

\item {\bf $qq$-characters.}  Generalized stress tensor operators
  $\hat W_a$ can be actually understood in terms of the DIM ${\cal
    R}$-matrices, and from this perspective they give abstract
  algebraic description of the $qq$-characters \cite{NPSh,Nekr,Pestun}. This
  construction generalizes ordinary $q$-characters for quantum groups
  introduced in \cite{FR}.

\item {\bf Systems of symmetric functions.}  One can associate with
  the set of matrix integrals/algebra a set of symmetric functions in
  two different ways. One option is to construct them directly from
  the integral, omitting one set of integrations \cite{AwataMac,Mac}.  In
  the simplest example of the matrix integral with $n=1$ we fix the
  Young diagram $\lambda$ with lengths of lines $\lambda_\alpha$,
  $\alpha=1\ldots (m-1)$ and the corresponding symmetric function of
  variables $x_i\equiv x_i^{(N)}$ corresponding to $\lambda$ is given
  by the matrix integral
  \be\label{sfun}
  P_{\lambda^{T}}
  (x_i)\sim \int\left[\prod_{\alpha=1}^{m-1}\prod_i^{\lambda_\alpha}dx_i^{(\alpha)}\right]\prod_{\alpha
    ,\beta=1}^{m}{\prod}_{i,j}^{\prime\lambda_\alpha,\lambda_\beta}
  \Delta\Big(\vec x_{\alpha},\vec x_{\beta}\Big)^{C_{\alpha\beta}}
  \ee
  where $\lambda_m$ is put equal to $m$. This is a generalization of old formulas from \cite{versus} and
and it can be considered as an extension of the degenerate field insertion into the conformal block \cite{MMMsurf} in the DF approach.

Another way to construct symmetric functions \cite{Mac,defMac}
is to consider a level 1 representation of the algebra so that it is
realized by the Heisenberg algebra.  Choose a Hamiltonian as an
element of the algebra, it is a function of generators $a_n$.
Realizing them in terms of time variables, $a_{n<0}\sim \ttau_n$,
$a_{n>0}\sim \partial_{\ttau_n}$, one obtains a set of eigenfunctions
of the Hamiltonian as functions of $\ttau_n$.  After the Miwa
transformation, $\ttau_n=\sum_i x_i^n$ they give rise to symmetric
functions of $x_i$.  For instance, the level one representation of the
DIM ${\mathfrak{gl}}_1$ algebra leads to the set of
${\mathfrak{gl}}_1$ Macdonald polynomials.  As a next step, one can
consider the sets of eigenfunctions which diagonalize co-products (of
degree $N-1$) of the Hamiltonian (i.e. representations of higher
levels), which, in this concrete example, leads to the generalized
${\mathfrak{gl}}_N$ Macdonald polynomials.

\item {\bf Lift to
the graph level.}
The next step is restoration of the vertical/horizontal symmetry and lifting
the symmetry (Ward identities) to the original network matrix model (\ref{netmod}).
Important at this step is that topological vertices
are associated with matrix elements of the intertwining operators of the DIM algebra \cite{AF}.
\end{itemize}

The crucial ingredient of this construction is the centralizer of the
algebra of constraints, defining the screening operators and the
matrix model.  The centralizer depends on the representation,
and it is this dependence that leads to a variety of different
matrix models, encoded by the graph $\Gamma$. Once the graph (with
some additional decorations: preferred direction, etc.) is chosen,
the particular representation of DIM is fixed and so is the
particular matrix model. However, traces of the larger DIM symmetry
remain in various forms, the most notable example being the
spectral duality~\cite{Bao,MMZ,spedu} connecting
multi-matrix models with different numbers of matrices and vertex
operator insertions.

\bigskip

In the simplest case, associated with $4d$ Seiberg-Witten theory, the
role of $\hat T$ is played by the stress tensor
$T(z)=\frac{1}{2}\p\phi(z)^2 + Q \partial^2 \phi(z)$, which generates
the ordinary Virasoro algebra (and its $W_N$-algebra generalizations),
and $Z(t)$ is just the ordinary Dotsenko-Fateev (DF) matrix model of
\cite{DF}.  Various types of $q/t/q_{123}$-deformations, associated
with reviving of the hidden compactification moduli, i.e. revealing
the hidden $6d$ and M-theory nature of the theory, require a
lifting/resolution of $\frac{1}{2}\p\phi(z)^2 + Q \partial^2 \phi(z)$
of a peculiar Toda-like combination of vertex operators:
\begin{equation}
{\cal T}(z) = \ : e^{\Phi(z)} e^{-\Phi(t^{-1}z)}: +\ t :
e^{-\Phi(tz/q)} e^{\Phi(z/q)}:
\label{Tdefo}
\end{equation}
where
\begin{equation}
  \label{eq:13}
  \Phi(z) = \sum_{n \geq 1} \frac{ z^n}{n} \alpha_{-n}
  + \Phi_0 - \sum_{n
  \geq 1} \frac{z^{-n}}{n} \alpha_n,
\end{equation}
and the modes of $\Phi(z)$ satisfy the $q$-deformed commutation
relations:
\begin{equation}
  \label{eq:14}
  [\alpha_n, \alpha_m] = \frac{n}{1 + \left( \frac{q}{t} \right)^{|n|}} \frac{1 - q^{|n|}}{1 - t^{|n|}}  \delta_{n+m,0}.
\end{equation}
Given these $q$-boson relations, $\mathcal{T}(z)$ generates the
$q$-deformed Virasoro algebra $\mathsf{Vir}_{q,t}$. Deformed
stress-energy tensor~\eqref{Tdefo} can be guessed from the requirement
that it commutes with the screening current. The expression for the
screening current essentially determines the matrix model and its
symmetry. Concretely, the $q$-deformed screening current is given by
\begin{equation}
  \label{eq:47}
  S(x) = \ : \prod_{k\geq 0}\exp \left(- \Phi(q^k x) + \Phi(q^k tx) + \Phi(
    q^{k+1} x) - \Phi \left( \frac{q^{k+1}}{t} x \right)\right) :
\end{equation}
We will derive the formulas for $\mathcal{T}$ and $S$ in detail in
sec.~\ref{sec:w-algebra-ward}, in particular we obtain the
form~\eqref{Tdefo} of stress-energy tensor in Eq.~\eqref{eq:37}.

One can see that the expressions for $\mathcal{T}$ and $S$ are not
symmetric under the exchange of $t$ and $q^{-1}$, which is, as we
will see, the natural symmetry of the DIM algebra. This is another
artifact of the choice of a concrete representation/matrix model
description of the object with larger symmetry. All the essential
quantities of each particular model should be symmetric w.r.t.\ $q
\leftrightarrow t^{-1}$, though the intermediate results do not
respect this symmetry.

In the double-scaling limit $q = e^{\hbar} \to 1$, $t = q^{\beta}$ the
ordinary Virasoro stress-energy tensor $T(z) = \frac{1}{2}\p\phi(z)^2
+ Q\p^2\phi(z)$ (with $Q = \sqrt{\beta} - \sqrt{\beta}^{-1}$) is
recovered from (\ref{Tdefo}):
\begin{equation}
  \label{eq:15}
  \mathcal{T}(z) \to 2 + \hbar (1-\beta)  + \frac{\hbar^2}{2} \left[
    (\beta - 1)^2 + z^2 T(z)  \right] + \mathcal{O}(\hbar^3),
\end{equation}
where
\begin{equation}
  \label{eq:16}
  \phi(z) = 2 \sqrt{\beta} \lim_{
    \begin{smallmatrix}
      q,t \to 1\\
t = q^{\beta}
    \end{smallmatrix}
  }  \Phi(z) = \sum_{n \geq 1} \frac{z^n}{n} \hat{\alpha}_{-n}
  + \Phi_0 - \sum_{n
    \geq 1} \frac{z^{-n}}{n} \hat{\alpha}_n,
\end{equation}
and $\hat{\alpha}_n$ are ordinary boson generators, satisfying
$[\hat{\alpha}_n , \hat{\alpha}_m] = 2 n \delta_{n+m,0}$.

As already mentioned,
a nice bonus is that multi-field generalization of
(\ref{Tdefo}), which in $4d$ leads to substitution of Virasoro by
$W$-algebras, is now just another representation of the {\it same}
symmetry algebra.  In other words, after the deformation the
Sugawara-like bi- and multi-linear combinations of currents can be
obtained from comultiplication of the deformed current algebra,
without a need to consider its universal enveloping.

\bigskip

The purpose of this paper is a sketchy survey of this remarkable DIM
symmetry of (\ref{netmod}).
Various details will be presented in separate publications.  We will
discuss here the $5d$ and $6d$ DIM $\mathfrak{gl}_1$ algebras which
correspond to the quiver gauge theories with fundamental matter. The
most interesting case of the DIM affine algebras which describe, in
particular, the $6d$ gauge theory with adjoint matter and correspond
to the double elliptic systems will be touched only briefly. This
issue, and also various details of other cases will be presented in
separate publications.

\section{$A_1$ $(q,t)$-matrix model}
\label{sec:a_1-q-t}

Let us start with the prototypical example of the $A_1$
$(q,t)$-deformed conformal matrix model. This is the simplest model
where $q$-Virasoro symmetry arises and, therefore, serves as an
accessible port of entry to the land of DIM algebras.

In this section we describe the general scheme for investigating a
network-type matrix model. We start by writing down the conventional
definition of the model in terms of matrix integral. However, one
should remember, that this is just a particular representation of the
network of topological vertices, as in Eq.~\eqref{top-network}. We
next describe the algebraic face of the matrix model more concretely
by specifying the screening operators, which OPE gives the actual
matrix model integrals. The \emph{centralizer} of the screenings
inside the representation of DIM gives the $W$-algebra corresponding
to the matrix model, which also generates the $qq$-characters in the
gauge theory. This description was used in~\cite{Pestun} to
introduce the $W$-algebras corresponding to an arbitrary (affine)
ADE-type quiver. Our aim in this paper is more general (though in this
section we study it on a very humble example). We would like to
elucidate the hidden symmetries, which are only visible in the
network-type formalism~\eqref{top-network} (see~\cite{MMZ} for an
example of such an approach). The symmetries of the
network/topological string/toric diagram are described by DIM algebra,
of which different $W$-algebras are only particular
representations/subalgebras. In this part of the paper we will
demonstrate explicitly how various concepts in matrix models and gauge
theories, such as $qq$-characters and generalized Macdonald
polynomials, are tied together with the help of the DIM algebra.

We introduce the DIM algebra generators and relations in
sec.~\ref{sec:dim-algebra}. We describe the simplest representations
of DIM algebra in sec.~\ref{sec:fock-repr-dim} and show how they give
rise to generalized Macdonald polynomials. In
sec.~\ref{sec:w-algebra-ward} with the help of dressing operators, we
build the deformed Virasoro subalgebra of the DIM algebra and show its
connection to $qq$-characters in the gauge theory. In
sec.~\ref{sec:mathsfvir_q-t-oplus} we focus on the details of the
dressing procedure and identify it with the reduction of the ``$U(1)$
part'' in the Nekrasov function/conformal block. We also describe the
relation with Benjamin-Ono integrable system.

\subsection{Free-field description}
\label{sec:free-field-descr}

The matrix model can be described in two different ways: as a
Jackson or contour integrals respectively. Here we adopt the latter
form:
\begin{equation}
  \label{eq:17}
  Z_{A_1} = \oint d^Nx\, \Delta^{(q,t)}(x) V_1(z_1, x) \cdots V_M(z_M, x)
  ,
\end{equation}
where
\begin{gather}
  \Delta^{(q,t)}(x) = \prod_{i \neq j} \frac{\left(\frac{x_i}{x_j};
      q\right)_{\infty} }{\left(t \frac{x_i}{x_j};
      q\right)_{\infty}},\notag\\
  V_a(z_a,x) = \prod_{i=1}^N \frac{\left( q^{1-v_a} \frac{z_a}{x_i}; q
    \right)_{\infty}}{\left( \frac{z_a}{x_i} ; q \right)_{\infty}}, \label{eq:18}
\end{gather}
and the Pochhammer symbol ($q$-exponential) is defined as $(x;
q)_{\infty} = \prod_{k \geq 0} (1 - q^k x)$. Time variables are traded
for a product of vertex operators $V(z)$: this can be
understood/interpreted as a Miwa transform.

Following the general recipe given in the introduction, we would like
to interpret the matrix model as an average of screening currents
$S(x)$. One can see explicitly that the necessary choice is
\begin{equation}
  \label{eq:19}
  S(x) =\ : \exp \left[- \sum_{n \geq 1} \frac{x^n}{n} \frac{1 - t^n}{1
      - q^n} \left( 1 + \left( \frac{q}{t} \right)^n \right)
    \alpha_{-n} + \sum_{n \geq 1} \frac{x^{-n}}{n} \frac{1 - t^{-n}}{1
      - q^{-n}} \left( 1 + \left( \frac{q}{t} \right)^{-n} \right)
    \alpha_n \right] :
\end{equation}
where $\Phi(x)$ is defined in Eq.~\eqref{eq:13}. From the $q$-boson
commutation relations~\eqref{eq:14} one get the following OPE for the
screenings currents
\begin{equation}
  \label{eq:21}
  S(x_1) S(x_2) =\ \frac{\left( \frac{x_2}{x_1}; q
    \right)_{\infty}\left( \frac{q}{t} \frac{x_2}{x_1}; q
    \right)_{\infty}}{\left( t \frac{x_2}{x_1}; q \right)_{\infty}
    \left( t \frac{x_2}{x_1}; q \right)_{\infty}} \ :S(x_1) S(x_2):
  \quad
\stackrel{\text{integer $\beta$}}{\longrightarrow}\ \  \prod_{k=0}^{\beta -1} \left( 1
    - q^k \frac{x_1}{x_2} \right) \left( 1
    - q^k \frac{x_2}{x_1} \right) :S(x_1) S(x_2):
\end{equation}
From the OPE~\eqref{eq:21} we can immediately see that the matrix
model~\eqref{eq:17} is indeed the correlator of screenings with vertex
operators:
\begin{equation}
  \label{eq:20}
  Z_{A_1} =\langle 0 | \oint d^Nx \prod_{i=1}^N S(x_i) V_1(z_1)
    \cdots V_M(z_M) | 0 \rangle
\end{equation}
What are the Ward identities for the $(q,t)$-matrix model? To obtain
them let us perform the steps we discussed in the Introduction: first, we introduce time variables $p_k$ into the matrix integral inserting into the average (\ref{eq:20}) the operator
\be\label{G}
{\cal G}(p)=\exp\Big(\sum_{k>0}p_{k}\alpha_{-k}\Big)
\ee
and, second, we verify that the deformed stress-energy tensor
$\mathcal{T}(z)$~\eqref{Tdefo} \emph{commutes} with the integral of
the screening current~\eqref{eq:19}, $\oint S(x) dx/x$. This means
that $\mathcal{T}(z)$ commutes with $S(x)$ up to total derivative (or
total $q$-difference). The OPE of $\mathcal{T}$ with $S$ is given by
\begin{multline}
  \label{eq:7}
  \mathcal{T}(z) S(x) = \frac{1 - t\frac{x}{z}}{1 - \frac{x}{z}}
  :e^{\Phi(z) - \Phi(z/t)} S(x): + t \frac{1 -
    \frac{q}{t}\frac{x}{z}}{1 - q\frac{x}{z}}
  :e^{-\Phi(tz/q) + \Phi(z/q)} S(x):=\\
  =\ :\Biggl\{ \left(\frac{1 - t \frac{x}{z}}{1 - \frac{x}{z}}e^{\Phi(z)-
    \Phi(x) + \Phi( x/t) -
    \Phi(z/t)  } + t \frac{1 - \frac{1}{t}\frac{x}{z}}{1 -
    \frac{x}{z}} e^{\Phi(t x/q) -\Phi(tz/q) + \Phi(z/q)
      -\Phi(x/q)} \right) e^{\Phi(x) - \Phi( x/t)}+ \\
  + (q^{x \partial_x} - 1) \left( t \frac{1 - \frac{1}{t}\frac{x}{z}}{1 -
    \frac{x}{z}} e^{\left[\Phi(t x/q) -\Phi(tz/q) + \Phi(z/q) -\Phi(x/q)\right]  + \Phi(x) - \Phi(
    x/t)}\right)  \Biggr\} S(x):
\end{multline}
Remarkably, the pole at $z = x$ is exactly canceled in both terms in
the first line: the shift in the infinite sum of operators inside
$S(x)$ plays a crucial role in this cancellation, and only the
total difference remains singular. Since the poles are canceled up to
total $q$-difference, the commutator with screening charge, i.e.\
with the integral of $S(x)$ vanishes. This fact was used
in~\cite{Pestun} to derive the regularity of the $qq$-characters.

This implies that $\mathcal{T}(z)$ is a symmetry of the model:
negative modes of its Laurent expansion in $z$ annihilate the vacuum
and thus annihilate the entire matrix integral.  Inserting the
$\mathcal{T}$-$S$ OPE into the matrix integral, we get:
\begin{multline}
\mathcal{T}_{p_k}(z)Z_{A_1}(p)=\Big\langle \mathcal{T}(z) \Big\rangle = \frac{\langle 0 |
  {\cal G}(p)  V_1(z_1)\cdots V_M(z_M) \mathcal{T}(z) \oint d^Nx \prod_{i=1}^N
    S(x_i) | 0 \rangle}{\langle 0 | V_1(z_1)\cdots V_M(z_M) \oint d^Nx
    \prod_{i=1}^N S(x_i)
    | 0 \rangle} =\\
  \oint d^Nx\, \Delta^{(q,t)}(x) \left(\prod_i V_1(z_1,x_i)\cdots V_M(z_M,x_i)
U(x_i,p)\right)
  \left( \prod_{i=1}^N \frac{1 -
      t \frac{x_i}{z}}{1 -  \frac{x_i}{z}} + P(z|\{z_a\},\{ v_a\})
    \prod_{i=1}^N \frac{1 - q
      \frac{x_i}{z}}{1 - \frac{q}{t} \frac{x_i}{z}} \right) =
  \mathrm{Pol}(z)
    \label{Tave}
\end{multline}
where $P(z|\{z_a\},\{ v_a\})$ is the contribution of vertex operators, a $z$-polynomial factor and the time dependence of the partition function is encoded in the potential
\be
U(x,p)=\exp \left(- \sum_{n \geq 1} \frac{x^n}{n} \frac{1 - t^n}{1
      - q^n} \left( 1 + \left( \frac{q}{t} \right)^n \right)
    p_n\right)
\ee
The deformed
stress-energy tensor, written in the bosonized form, as in
Eq.~\eqref{Tdefo}, or in the form of matrix model average, can also be realized as a difference operator upon identification
\be
\hat\alpha_{-n}=p_n,\ \ \ \ \ \ \hat\alpha_n=\frac{n}{1 + \left( \frac{q}{t} \right)^{n}} \frac{1 - q^{n}}{1 - t^{n}} {\partial\over \partial p_n}
\ee
leading to a difference equation on the partition function, a counterpart of the Baxter equation.
It is sometime called a $qq$-character~\cite{NPSh,Pestun,Nekr,Mat}, since it can be considered
as a deformation of the Frenkel-Reshetikhin $q$-character \cite{FR}
(trace over Cartan part of the quantum ${\cal R}$-matrix).
Virasoro symmetry of the matrix model implies that this average
has no negative modes in its $z$-expansion, i.e.\ is regular (and
therefore polynomial) in $z$:
\begin{multline}
  \label{eq:5}
  \text{regularity of the} \ qq\text{-character} =
  \text{polynomiality of the average} \ \langle \mathcal{T}(z)\rangle \ =\\
  = \text{Ward identity (DIM/Virasoro constraint)}
\end{multline}
While obviously following from commutativity of $\mathcal{T}(z)$ with
$S$, this looks like a non-trivial property of the r.h.s. in
\eqref{Tave}.

Also $qq$-characters can be thought of as the recurrence relation on
the matrix model correlators, obtained by expanding the average of
$\mathcal{T}(z)$ in powers of $z$. The recurrence relations can also
be derived by considering the vanishing total difference under the
matrix model integral~\cite{MMZ}. Of course, this only means that
the commutator of $\mathcal{T}(z)$ and $S(x)$ is given by the
corresponding total difference. In the case at hand the relevant total
difference is given by
\begin{multline}
  \label{eq:48}
  0 = \oint d^N x \sum_{i=1}^N \frac{1}{x_i} \left( 1 -
    q^{x_i \partial_i} \right) \left[ \frac{x_i}{z - x_i} \prod_{j
      \neq i} \frac{x_i
      - t x_j}{ x_i - x_j} \Delta^{(q,t)}(x) \right] =\\
  = \oint d^N x \sum_{i=1}^N \left[ \frac{1}{z - x_i} \prod_{j \neq i}
    \frac{x_i - t x_j}{ x_i - x_j} - \frac{t^{N-1} q}{z - q x_i}
    \prod_{j \neq i} \frac{ t x_i - x_j}{ x_i - x_j} \right]
  \Delta^{(q,t)}(x) =
  \\
  = \oint d^Nx \left[ \prod_{j=1}^N \frac{1 - \frac{t x_j}{z}}{1 - \frac{x_j}{z}} +
    t^{2N-1} q \prod_{j=1}^N \frac{1 - \frac{q x_j}{tz}}{ 1 - \frac{q x_j}{z}} - Q_N(z) \right]
\end{multline}
where $Q_N(z)$ is degree $N$ polynomial in $z$ and $x_i$, and in the
last line we have summed over poles in $z$ to obtain the products. The
identity~\eqref{eq:48} is precisely the regularity constraint on the
$qq$-character telling that $\langle \mathcal{T}(z)\rangle = \langle
P(z)\rangle = \text{regular in $z$}$. For details of derivation along
this route see~\cite{MMZ}. We will employ similar technique to get
the symmetry constraints for the elliptic matrix model in
sec.~\ref{sec:elliptic-dim-algebra}.

In the next section we show how to obtain the deformed energy-momentum
tensor from the representation of the abstract DIM algebra.

\subsection{Abstract algebraic description}
\label{sec:abstr-algebr-descr}

We now describe the algebraic structures of DIM algebra governing the
network-type matrix model. Let us first recall the definition of the
DIM algebra $U_q (\widehat{\widehat{\mathfrak{gl}}}_1)$ and its
simplest representations and then demonstrate the connections of this
algebra with deformed Virasoro algebra, $qq$-characters, generalized
Macdonald polynomials and integrable systems.

\subsubsection{DIM algebra}
\label{sec:dim-algebra}
This looks like a deformation of the affine quantum algebra
$U_q(\widehat{\mathfrak{gl}}_2)$ with the positive/negative root
generators $x^\pm(z)$, two exponentiated Cartan generators
$\psi^\pm(z)$ and the central element $\gamma$.

Commutation relations are
\begin{gather}
  G^{\mp}(z/w)\,x^\pm(z)\,x^\pm(w) \ =\  G^\pm(z/w)\, x^\pm(w)\, x^\pm(z)\nn \\
  \l[x^+(z),\, x^-(w)]\ =\
  \frac{(1-q)(1-t^{-1})}{1-q/t}\,\Big(\delta(\gamma^{-1}z/w)\,\psi^+(\gamma^{1/2}w)
  \ -\ \delta(\gamma z/w)\,\psi^-(\gamma^{-1/2}w)\Big)\nn \\
  \psi^\pm(z)\,\psi^\pm(w) \ = \ \psi^\pm(w)\,\psi^\pm(z) \label{eq:24} \\
  \psi^+(z)\,\psi^-(w)\ = \ \frac{g(\gamma w/z)}{g(\gamma^{-1}w/z)}\, \psi^-(w)\psi^+(z) \nn \\
  \psi^+(z)\, x^\pm(w) \ =\  g(\gamma^{\mp 1/2}w/z)^{\mp 1}\, x^\pm(w)\,\psi^+(z) \nn \\
  \psi^-(z)\, x^\pm(w) \ =\ g(\gamma^{\mp 1/2}z/w)^{\pm 1}\,
  x^\pm(w)\,\psi^-(z)\nn\\
  \mathop{\mathrm{Sym}}\limits_{z_1, z_2, z_3} z_2 z_3^{-1} [x^{\pm}(z_1),
  [x^{\pm}(z_2),x^{\pm}(z_3)]]=0\nn
\end{gather}
DIM algebra is a Hopf algebra with comultiplication
\begin{gather}
  \Delta\Big(\psi^\pm(z)\Big) \ = \ \psi^\pm(\gamma^{\pm 1/2}_2z)\,\otimes\, \psi^\pm(\gamma^{\mp 1/2}_1z)\nn\\
  \Delta\Big(x^+(z)\Big)\ = \ \psi^-(\gamma^{ 1/2}_1
  z)\,\otimes\,x^+(\gamma_1 z) \ + \ x^+(z)\,\otimes\, 1 \label{eq:23} \\
  \Delta\Big(x^-(z)\Big)\ = \ 1\,\otimes\, x^-(z)\ + \ x^-(\gamma_2
  z)\,\otimes\,\psi^+(\gamma^{ 1/2}_2 z)\nn
\end{gather}
where $\ \gamma_1^{\pm 1/2} = \gamma^{\pm 1/2}\otimes 1, \ \ \ \
\gamma_2^{\pm 1/2} = 1\otimes \gamma^{\pm 1/2}$. The function $g(z) =
\frac{G^+(z)}{G^-(z)}$ is restricted by the requirement
$g(z)=g(z^{-1})^{-1}$ and $\delta(z) = \sum_{n\in \mathbb{Z}}z^n$. We
omit expression for the counit and antipode, since we will not need
them.

\subsubsection{Specification of the structure function}

The structure of the algebra is encoded in the function $G(z)$
which is often chosen to be \emph{cubic} in $z$ with additional
restriction $q_1q_2q_3=1$:
\begin{equation}
G^\pm(z) = (1 - q_1 z) (1 - q_2 z) (1 - q_3 z)  = \left(1-q^{\pm 1}z\right)\left(1-t^{\mp 1}z\right)\left(1-(t/q)^{\pm 1}z\right)
\end{equation}
Without any harm to commutation relations and comultiplication
it can be further promoted to
unrestricted $q_{123}$- and more general Kerov deformations, and even
to elliptic function, though details of bosonization procedure below should still be
worked out in these cases.  We describe the elliptic version
in sec.~\ref{sec:elliptic-dim-algebra}.

\subsubsection{Level one Fock representation}
\label{sec:fock-repr-dim}
The simplest representation of DIM algebra is the level one
representation $\rho_u$ acting on the Fock module $\mathcal{F}_u$,
generated by the $q$-deformed Heisenberg creation
operators $a_{-n}$ from the vacuum $|u \rangle$ annihilated by the annihilation
operators $a_n$. The Heisenberg generators satisfy
\begin{equation}
  \label{eq:46}
  [a_n, a_m] = n \frac{1 - q^{|n|}}{1 - t^{|n|}} \delta_{n+m,0}
\end{equation}
Note that $a_n$ are normalized differently from $\alpha_n$ in
eqs.~\eqref{eq:13},~\eqref{eq:14} (that normalization was chosen to
maximally simplify the final expressions). Of course the $a_n$
generators are related to $\alpha_n$ generators in a simple way:
\begin{gather}
  \label{eq:1}
  \alpha_n = \frac{1}{1 + (q/t)^n} a_n \quad n \geq 1\\
  \alpha_{-n} = a_{-n} \quad n \geq 1
\end{gather}

The generators of the DIM algebra are expressed in terms of the
Heisenberg generators:
\begin{gather}
  \rho_u\left(x^+(z)\right) = u \eta(z) = u\ :\exp \left( \sum_{n \geq
      1} \frac{1 - t^{-n}}{n} a_{-n} z^n - \sum_{n \geq 1} \frac{1 -
      t^n}{n} a_n z^{-n} \right): \nn \\
  \rho_u\left(x^-(z)\right) = u^{-1} \xi(z) = u^{-1}\ :\exp \left(
    \sum_{n \geq 1} \frac{1 - t^{-n}}{n} \left( \frac{t}{q}
    \right)^{n/2} a_{-n} z^n - \sum_{n \geq 1} \frac{1 -
      t^n}{n} \left( \frac{t}{q} \right)^{n/2} a_n z^{-n} \right):  \nn \\
  \rho_u\left(\psi^\pm(z)\right) = \varphi_\pm (z) = \exp \left( \mp
    \sum_{n \geq 1} \frac{1 - t^{\pm n}}{n} \left( 1 - \left(
        \frac{t}{q} \right)^n \right)a_{\pm n} z^{\mp n} \right)\label{eq:40}\\
  \rho_u(\gamma) = \left( \frac{t}{q} \right)^{1/2} \nn
\end{gather}
Let us see an example how OPE of these operators reproduces the DIM
commutation relation:
\begin{equation}
  \label{eq:6}
  \eta (z) \eta (y) = \frac{\left(1 - \frac{z}{y}\right) \left( 1 -
      \frac{q}{t} \frac{z}{y} \right)}{\left(1 - \frac{1}{t} \frac{z}{y}\right) \left( 1 -
      q \frac{z}{y} \right)} \ : \eta (z) \eta (y): = \frac{\left( 1 -
      \frac{q}{t} \frac{z}{y} \right) \left(1 - \frac{1}{t} \frac{y}{z}\right) \left( 1 -
      q \frac{y}{z} \right)}{\left(1 - \frac{1}{t} \frac{z}{y}\right) \left( 1 -
      q \frac{z}{y} \right) \left( 1 -
      \frac{q}{t} \frac{y}{z} \right)} \eta (z) \eta (y) = \frac{G^{-}
    \left( \frac{z}{y} \right)}{G^{+} \left( \frac{z}{y} \right)} \eta (z) \eta (y)
\end{equation}

\subsubsection{Level two Fock representation and generalized Macdonald
  polynomials}
\label{sec:level-two-fock}
Tensor product of $m$ Fock representations $\mathcal{F}_{u_1} \otimes
\cdots \otimes \mathcal{F}_{u_m}$ can be easily obtained from the
comultiplication~\eqref{eq:23} and will be called the level $m$ Fock
representation. In this tensor product the generators of DIM algebra
are expressed in terms of $m$ $q$-Heisenberg generators $a_n^{(a)}$,
$a =1, \ldots ,m$. In particular, we will need the expression for
$x^{+}(z)$ in this representation:
\begin{equation}
  \label{eq:25}
  \rho_{u_1,u_2}^{(2)} (x^{+}(z)) = u_1
  \widetilde{\Lambda}_1 (z) + u_2 \widetilde{\Lambda}_2(z) = u_1 \eta_1
  (z)   + u_2 \varphi_1^{-}\left( \left(t/q \right)^{1/4} z
  \right) \eta_2 \left( \left( t/q \right)^{1/2} z  \right).
\end{equation}
where we use the shorthand notation $\widetilde{\Lambda}_{1,2}$ for
the components of the level two representation $\rho_{u_1,u_2}^{(2)} =
(\rho_{u_1} \otimes \rho_{u_2}) \Delta$, and the subscript denotes the
number of term in the tensor product, e.g.\ $\eta_1(z) =
\eta(z)\otimes 1$.

There is an distinguished basis in $\mathcal{F}_{u_1} \otimes \cdots
\otimes \mathcal{F}_{u_m}$, the basis of generalized Macdonald
polynomials \cite{defMac} obtained by diagonalizing
the action of the zero mode of $x^{+}(z)$. Representation of this zero
mode was called generalized Macdonald Hamiltonian in:
\begin{equation}
  \label{eq:30}
  H_1^{\mathrm{gen}} = \rho_{u_1,u_2}^{(2)} ( x^{+}_0) = \oint_{\mathcal{C}_0} \frac{dz}{z}\rho_{u_1,u_2}^{(2)}( x^{+}(z)),
\end{equation}
In those papers the following definition of the generalized Macdonald
polynomials was given:
\begin{equation}
  \label{eq:31}
  H_1^{\mathrm{gen}} M_{AB}(a^{(1)}_{-n}, a^{(2)}_{-n}) |u_1 \otimes
  u_2 \rangle = \left[u_1\kappa_A(q,t) + u_2\kappa_B(q,t)\right] M_{AB}(a^{(1)}_{-n}, a^{(2)}_{-n}) |u_1 \otimes
  u_2 \rangle,
\end{equation}
where
\begin{equation}
  \label{eq:32}
  \kappa_A = (1-t) \sum_{i \geq 1} q^{A_i} t^{-i}.
\end{equation}
These polynomials were instrumental in demonstrating the $5d$ version
of the AGT conjecture \cite{5dAGT}. Matrix elements of Virasoro primary fields in
this basis turned out to coincide with fixed point contributions in
the Nekrasov partition function. Thus, after decomposition of
conformal blocks in terms of generalized Macdonald polynomials, the
AGT relation becomes explicit. In the $4d$ limit this special basis
degenerates into the basis of generalized Jack polynomials
\cite{defMac}, with similar properties.

\subsubsection{$W$-algebra, Ward identities and $qq$-characters from
  DIM}
\label{sec:w-algebra-ward}
As we have announced in the introduction, the great benefit of DIM
approach is that it describes different matrix models from a unified
viewpoint. In particular, $m$-multimatrix models have $W_m$-algebra
symmetries, and these algebras are all particular representations of
subalgebras of DIM algebra.

$q$-deformed $W_m$-algebra, which is also called $W_{q,t}
(\mathfrak{sl}_m)$, is obtained from level $m$ Fock representation of
the DIM algebra as follows. The stress-energy tensor of the
$W_m$-algebra is obtained from the \emph{dressing} of the $x^{+}$
generator of DIM. More concretely, we have:
\begin{equation}
  \label{eq:26}
  t(z) = A (z) x^{+}(z) B(z),
\end{equation}
where
\begin{equation}
  \label{eq:27}
  A(z) = \exp \left(  - \sum_{n \geq 1} \frac{1}{\gamma^n -
      \gamma^{-n}} b_{-n} z^n \right), \qquad B(z) = \exp \left( \sum_{n \geq 1} \frac{1}{\gamma^n -
      \gamma^{-n}} b_n z^{-n} \right)
\end{equation}
and $b_n$ are the modes of the $\psi^{\pm}$ generators:
\begin{equation}
  \label{eq:28}
  \psi^{\pm}(z) = \psi_0^{\pm} \exp \left( \pm \sum_{n \geq 1} b_{\pm
      n} \gamma^{n/2} z^{\mp n} \right).
\end{equation}
The stress-energy tensor $\mathcal{T}$ of the $W_M$-algebra is the
representation of the dressed current $t(z)$ in the level $m$ Fock
module. For the Virasoro case ($m=2$), using Eq.~\eqref{eq:25}, we get
\begin{multline}
  \label{eq:29}
  \mathcal{T}(z) = \rho_{u_1,u_2}^{(2)} (t(z)) = u_1
  \Lambda_1(z) + u_2 \Lambda_2(z)=\\
  = \rho_{u_1, u_2}^{(2)}(A(z)) \left( u_1
    \eta_1 (z) + u_2 \varphi_1^{-}\left( \left(t/q \right)^{1/4} z
    \right) \eta_2 \left( \left( t/q \right)^{1/2} z \right) \right)
  \rho_{u_1, u_2}^{(2)}(B(z)).
\end{multline}
where $\Lambda_i(z)$ are dressed versions of the components
$\widetilde{L}_i(z)$. From Eq.~\eqref{eq:29} we see that
$\mathcal{T}(z)$ depends on \emph{two} sets of Heisenberg generators
(hidden inside $\eta_1$, $\eta_2$ and $\varphi^{-}$) acting on the
tensor product of two Fock modules. However, as we will see explicitly
in the next section, the expression for $\mathcal{T}$ actually depends
only on \emph{one} linear combination of $a_n^{(1)}$ and
$a_n^{(2)}$. Related to this fact is that in the level two
representation the product of $\Lambda_{1,2}$ elements is equal to
identity:
\begin{equation}
  \label{eq:34}
  : \Lambda_1(z) \Lambda_2 \left( z q/t  \right):\ = 1.
\end{equation}
To see this fact we should write explicit (though lengthy) expressions
for $\Lambda_{1,2}$ in the level two representation:
\begin{gather}
  \label{eq:33}
  \Lambda_1(z) =\ : \exp \left( \sum_{n \geq 1} \frac{1}{n} \frac{1 -
      t^{-n}}{1 + \left( q/t \right)^n} z^n \left( \alpha_{-n}^{(1)} -
      \left( q/t \right)^{n/2} \alpha_{-n}^{(2)} \right)  - \sum_{n \geq 1} \frac{1 - t^n}{n} z^{-n} \left(
      \alpha_n^{(1)} - \left( q/t
      \right)^{n/2} \alpha_n^{(2)} \right) \right):\\
  \Lambda_2(z) =\ : \exp \left( - \sum_{n \geq 1} \frac{1}{n} \frac{1 -
      t^{-n}}{1 + \left( q/t \right)^n} \left( z t/q \right)^n
    \left( \alpha_{-n}^{(1)} - \left( q/t \right)^{n/2} \alpha_{-n}^{(2)}
    \right) + \sum_{n \geq 1} \frac{1 - t^n}{n}
    \left(z^{-1} q/t \right)^n \left( \alpha_n^{(1)} - \left( q/t
      \right)^{n/2} \alpha_n^{(2)} \right) \right):
\end{gather}
From these expressions we see that indeed $:\Lambda_1(z)
\Lambda_2 \left( z q/t \right) :\ = 1$. We also identify the
combinations of creation and annihilation operators, on which
$\mathcal{T}$ depends, and denote these combinations by
$\widetilde{\alpha}$. They are given by
\begin{align}
  \label{eq:35}
  \widetilde{\alpha}_{-n} &= \frac{1}{1 + \left( q/t \right)^n}
  \left(\alpha^{(1)}_{-n} - \left( q/t \right)^{n/2}
    \alpha^{(2)}_{-n}\right), \quad n \geq 1\\
  \widetilde{\alpha}_n &= \left(\alpha^{(1)}_n - \left( q/t
    \right)^{n/2} \alpha^{(2)}_n\right), \quad n \geq
  1 \label{eq:36}
\end{align}
One can see that the commutation relations for $\widetilde{\alpha}_n$
are the same as for $\alpha_n^{(1)}$. Now $\Lambda_{1,2}$ and the
stress-energy tensor $\mathcal{T}$ are all nicely written in terms of
these combinations:
\begin{equation}
  \label{eq:37}
  \mathcal{T}(z) = u_1 : e^{\widetilde{\Phi}(z)}
  e^{-\widetilde{\Phi}(t^{-1}z)}: + u_2 :
  e^{-\widetilde{\Phi}(tz/q)} e^{\widetilde{\Phi}(z/q)}:
\end{equation}
where the definition of $\widetilde{\Phi}$ is similar to that of
$\Phi$ from Eq.~\eqref{eq:13}, only the role of bosons $\alpha_n$ is
now played by $\widetilde{\alpha}_n$. The constants $u_1$ and $u_{2}$
can be absorbed into the definition of zero modes, which brings
Eq.~\eqref{eq:37} into the form of the deformed stress-energy tensor
identity~\eqref{Tdefo}. The zero modes of the screening operators are
omitted to simplify the formulas.

Finally, the operators $\widetilde{\alpha}_n$ are in fact precisely
those bosonic operators, in terms of which we have defined our matrix
model~\eqref{eq:20}. We have, therefore, identified the Ward
identities/Virasoro constraints of the matrix model with the
particular combination of the DIM operators in the level two Fock
representation. We can also make the identification with
$qq$-character more explicit by introducing the usual notation:
\begin{equation}
  \label{eq:38}
  \Lambda_1(z) = \mathcal{Y}(z),\qquad
  \Lambda_2(z) = \mathcal{Y}^{-1}\left( \frac{t}{q} z \right).
\end{equation}
The last definition follows from the condition~\eqref{eq:34}. Now we
would like to understand where the \emph{other} combination of the
bosonic generators is hidden. To see this we have to revisit the
dressing procedure, for the current $t(z)$.

\subsubsection{$\mathsf{Vir}_{q,t} \oplus \mathsf{Heis}_{q,t}$ reduction is
  equivalent to dressing}
\label{sec:mathsfvir_q-t-oplus}
In this section we show how the dressing operators $\alpha(z)$ and
$\beta(z)$ are in fact performing the reduction of the algebra
$\mathsf{Vir}_{q,t} \oplus \mathsf{Heis}_{q,t}$ acting in the level
two Fock representation to its $\mathsf{Vir}_{q,t}$ part. The
condition~\eqref{eq:34} can be thought of as a gauge condition used to
kill the $\mathsf{Heis}_{q,t}$ degrees of freedom, which enter both
$\Lambda_1$ and $\Lambda_2$ multiplicatively. This separation of
variables is usual for description of Hamiltonian reductions in the
free field formalism \cite{GMMold}.

To this end let us look at the bosonization of the dressing operators
$A(z)$ and $B(z)$. From Eqs.~\eqref{eq:40},~\eqref{eq:27} we
get
\begin{gather}
  \rho^{(2)}_{u_1, u_2}(A(z)) = \exp \left( - \sum_{n \geq 1}
    \frac{1}{n} \frac{1 - t^{-n}}{1 + \left( q/t \right)^n} \left(q/t
    \right)^{n/2} z^n \left( \left( q/t \right)^{n/2}
      \alpha^{(1)}_{-n} + \alpha_{-n}^{(2)} \right)
  \right) \label{eq:39},\\
  \rho^{(2)}_{u_1, u_2}(B(z)) = \exp \left( \sum_{n \geq 1}
     \frac{1 - t^n}{n} \left(q/t
    \right)^{n/2} z^{-n} \left( \left( q/t \right)^{n/2}
      \alpha^{(1)}_n + \alpha_n^{(2)} \right)
  \right).\notag
\end{gather}
In the exponent, these two operators contain precisely the linear
combination of $\alpha_n^{(1,2)}$ orthogonal to
$\widetilde{\alpha}_n$. We denote the new bosons by $\bar{\alpha}_n$:
\begin{align}
  \bar{\alpha}_{-n} = \frac{\left( q/t \right)^{n/2}}{1 + \left( q/t
    \right)^{n}} \left( \left( q/t \right)^{n/2} \alpha_{-n}^{(1)} +
    \alpha_{-n}^{(2)} \right),\label{eq:41}\\
  \bar{\alpha}_n = \left( q/t \right)^{n/2} \left( \left( q/t
    \right)^{n/2} \alpha_n^{(1)} +
     \alpha_n^{(2)} \right).\notag
\end{align}
These bosons commute with $\widetilde{\alpha}_n$ and satisfy slightly
modified (compared to~\eqref{eq:14}) commutation relations among
themselves:
\begin{equation}
  \label{eq:42}
  [\bar{\alpha}_n , \bar{\alpha}_m] = \frac{n (1-q^n) (q/t)^n}{\left( 1 +
      (q/t)^n \right) (1-t^n)} \delta_{n+m, 0}.
\end{equation}
Since the $\mathsf{Vir}_{q,t}$ algebra is entirely built out of
$\widetilde{\alpha}_n$, the new generators $\bar{\alpha}_n$ commute
with $\mathsf{Vir}_{q,t}$ and form the additional $q$-deformed
Heisenberg algebra $\mathsf{Heis}$. One can recall that such a
situation is common in the study of AGT relations \cite{AGT}, where Nekrasov
function \cite{LMNS,Nekrasov} usually corresponds to the conformal block \cite{CFT} of the Virasoro
algebra times an additional ``$U(1)$ factor'', which corresponds to an
extra boson, forming the $\mathsf{Heis}$ algebra \cite{Belavin}. Here we get the
extra boson for similar reasons: we are working in the tensor product
of two Fock modules, and have to eliminate the ``diagonal part'' of
the bosonized algebra. This elimination corresponds to the dressing
transformation, which is nothing but the transformation to the
``center of mass frame'' for the two bosons $\alpha^{(1,2)}_n$.

Finally, we can write down a compact expression for the
\emph{undressed} current $x^{+}(z)$ in the level two Fock
representation:
\begin{equation}
  \label{eq:43}
  \rho^{(2)}_{u_1, u_2} \left(x^{+}(z)\right) = u_1
  \widetilde{\Lambda}_1(z) + u_2
  \widetilde{\Lambda}_2(z)
  = \mathcal{T}(z) \mathcal{Z}(z) = \mathcal{T}(z)\ : e^{\bar{\Phi}(z)
    - \bar{\Phi}(z/t)}:
\end{equation}
where $\bar{\Phi}(z)$ is again the bosonic field defined analogously
to~\eqref{eq:13} using $\bar{\alpha}_n$ generators. We introduced the
$\mathsf{Heis}_{q,t}$ $qq$-character $\mathcal{Z}(z)$, in terms of
which the undressed current factorizes into the product of two terms
corresponding to algebras in $\mathsf{Vir}_{q,t} \oplus
\mathsf{Heis}_{q,t}$.

The factorized form of the current $x^{+}(z)$ is also reflected in
structure of its zero mode: the $H_1^{\mathrm{gen}}$
operator. Written in this form it gives the trigonometric
generalization of the Benjamin-Ono (BO)
equation~\cite{FL}, the continuous integrable model
also related to the AGT correspondence. It is easy to see the
structure of the BO Hamiltonians in the double scaling limit $q \to
1$, $t = q^{\beta}$:
\begin{equation}
  \label{eq:44}
  \oint_{\mathcal{C}_0} \frac{dz}{z} \rho^{(2)}_{u_1, u_2}(x^{+}(z)) =
  2 + \hbar (1 - \beta) + \frac{\hbar^2}{2} ( \mathbf{I}_1 + C_1) +
  \frac{\hbar^3 \beta}{2} (\mathbf{I}_2 + C_2) + \mathcal{O}(\hbar^4),
\end{equation}
where $C_{1,2}$ are constants,
\begin{align}
  \mathbf{I}_1 &= L_0 + 2\sum_{n\geq 1} \hat{\bar{\alpha}}_{-n}
  \hat{\bar{\alpha}}_n - \frac{1 - 3 Q^2}{6},\label{eq:45}\\
  \mathbf{I}_2 &= \sum_{k \neq 0} \hat{\bar{\alpha}}_{-k} L_k + 2 Q
  \sum_{n\geq 1} n \hat{\bar{\alpha}}_{-n} \hat{\bar{\alpha}}_n +
  \frac{1}{3} \sum_{n+m+k=0} \hat{\bar{\alpha}}_n \hat{\bar{\alpha}}_m
  \hat{\bar{\alpha}}_k,
\end{align}
and $\hat{\bar{\alpha}}_n$ are the ordinary Heisenberg generators,
obtained from $\bar{\alpha}_n$ in the double scaling limit. All higher
BO Hamiltonians appear in the higher terms. What we have found is that
generalized Macdonald polynomials are in fact joint polynomial
eigenfunctions of the quantum BO system.

\section{Elliptic DIM algebra and elliptic matrix model}
\label{sec:elliptic-dim-algebra}

In this section we describe the elliptic generalization of the matrix
model and DIM algebra governing it. As we will see, most of the
discussion is exactly parallel to the trigonometric case. This is
another manifestation of the universality of network type matrix
models and the DIM algebra. The fact that the description of the
elliptic case is so similar to the trigonometric one gives one the
hope that the corresponding structure in the double elliptic case
might also be tractable.

\subsection{Elliptic matrix model}
\label{sec:ellipt-matr-model}

This matrix model has been described in~\cite{Iq,ellMMZ}, and we
follow the notations of this paper.
\begin{equation}
  \label{eq:3}
  Z_{A_1}^{\mathrm{ell}} = \oint d^Nx \,
  \Delta^{(q,q',t)}_{\mathrm{ell}}(x) V_1(z_1, x) \cdots V_M(z_M, x),
\end{equation}
where
\begin{gather}
  \label{eq:4}
  \Delta_{\mathrm{ell}}^{(q,q',t)}(x) =\prod_{i \neq j} \frac{\left(
      \frac{x_i}{x_j} ; q, q' \right)_{\infty}\left( \frac{q q'}{t}
      \frac{x_j}{x_i} ; q, q' \right)_{\infty}}{\left( t
      \frac{x_i}{x_j} ; q, q' \right)_{\infty}\left( q q'
      \frac{x_j}{x_i} ; q, q' \right)_{\infty}},\\
  V_a(z_a, x) = \prod_{i=1}^N \frac{ \left( q^{1-v} \frac{z}{x_i} ;
      q,q' \right) \left( q q' \frac{x_i}{z} ; q,q' \right)}{\left(
       \frac{z}{x_i} ; q,q' \right) \left( q' q^v \frac{x_i}{z} ; q,q' \right)},
\end{gather}
where the double $q$-Pochhammer symbol is $(z;q,q')_{\infty} =
\prod_{k,l \geq 0} (1 - z q^k q'^l)$.

This elliptic integral arises from the following screening currents:
\begin{multline}
  \label{eq:9}
  S(x) =\ : \prod_{k\geq 0}\exp \left(- \hat{\Phi}(q^k x) + \hat{\Phi}(q^k tx) + \hat{\Phi}(
    q^{k+1} x) - \hat{\Phi} \left( \frac{q^{k+1}}{t} x \right)\right)
  :=\\
  =\ : \exp \left[- \sum_{n \neq 0} \frac{x^n}{n} \frac{1 - t^n}{(1
      - q^n)(1 - q'^{|n|})} \left( 1 + \left( \frac{q}{t} \right)^n \right)
    \hat{\alpha}_{-n} + \sum_{n \neq 0} \frac{x^{-n}}{n} \frac{1 - t^n}{(1
      - q^n)(1 - q'^{|n|})} \left( 1 + \left( \frac{q}{t} \right)^n \right)
    \hat{\beta}_{-n} \right] :
\end{multline}
where the bosons $\hat{\alpha}_n$ and $\hat{\beta}_n$ obey the
commutation relations:
\begin{gather}
    [\hat{\alpha}_n, \hat{\alpha}_m] = \frac{n (1 - q'^{|n|}) }{1 +
      \left( \frac{q}{t} \right)^{|n|}} \frac{1 - q^{|n|}}{1 -
      t^{|n|}}  \delta_{n+m,0},\notag\\
        [\hat{\beta}_n, \hat{\beta}_m] = \frac{n q'^{|n|} (1 -
          q'^{|n|}) }{1 + \left( \frac{q}{t} \right)^{|n|}} \frac{1 -
          q^{|n|}}{1 - t^{|n|}}  \delta_{n+m,0},  \label{eq:10}\\
        [\hat{\alpha}_n, \hat{\beta}_m] = 0.
\end{gather}
and $\hat{\Phi}(z)$ is the field built out of $\hat{\alpha}_n$ and
$\hat{\beta}_n$:
\begin{equation}
  \label{eq:11}
  \hat{\Phi}(z) = \sum_{n \neq 0} \frac{z^n}{n (1 - q'^{|n|})}
  \hat{\alpha}_{-n} - \sum_{n \neq 0} \frac{z^{-n}}{n (1 - q'^{|n|})} \hat{\beta}_{-n}
\end{equation}

Notice the presence of \emph{two} sets of boson generators
$\hat{\alpha}_n$ and $\hat{\beta}_n$, which is related to the modular
invariance of the elliptic model. More concretely two bosons produce
two terms in the product representation the theta-function: $\prod_{k
  \geq 0} (1 - q'^kz)(1 - q'^{k} q'/z)$, the elliptic version of
the free field correlator $(1-z)$. This explains why the powers of $z$
in front of $\hat{\alpha}_n$ and $\hat{\beta}_n$ are opposite, and
also why their commutation relation differ by $q'^{|n|}$.

Of course, the stress-energy tensor, which generates the centralizer
of the screening charge $Q = \oint S(x) dx/x$ also depends on two sets
of bosonic variables. It is very analogous to the trigonometric case:
\begin{equation}
  \label{eq:12}
  \mathcal{T}(z) =\ : e^{\hat{\Phi}(z)} e^{-\hat{\Phi}(t^{-1}z)}: +\ t :
e^{-\hat{\Phi}(tz/q)} e^{\hat{\Phi}(z/q)}:
\end{equation}
This elliptic stress-energy tensor generates the elliptic deformation
of the Virasoro algebra, which has been considered in many
works~\cite{Iq}. We proceed along the lines of the
previous section and move to the corresponding DIM algebra, which
gives tensor $\mathcal{T}$ in the level two representation.

\subsection{Elliptic DIM algebra, elliptic Virasoro and ILW equation}
\label{sec:dim-algebra-1}

Elliptic version of DIM algebra is generated by the same set of
operators as the ordinary DIM: $x^{\pm}(z)$, $\psi^{\pm}(z)$ and the
central element $\gamma$. The relations are a copy of
Eq.~\eqref{eq:24}, except for the $[x^{+}, x^{-}]$ relation, which
changes to
\begin{equation}
  \label{eq:2}
  \l[x^+(z),\, x^-(w)]\ =\
  \frac{\Theta_{q'} (q; q') \Theta_{q'} (t^{-1};
    q')}{(q';q')_{\infty}^3 \Theta_{q'} (q/t ; q')}\,\Big(\delta(\gamma^{-1}z/w)\,\psi^+(\gamma^{1/2}w)
  \ -\ \delta(\gamma z/w)\,\psi^-(\gamma^{-1/2}w)\Big)
\end{equation}
where $\Theta_p(z) = (p;p)_{\infty} (z;p)_{\infty} (p/z;p)_{\infty}$
is the theta-function. Also, most importantly, the structure function
$G^{\pm}(z)$ is now not trigonometric, but elliptic:
\begin{equation}
  \label{eq:49}
  G^{\pm}_{\mathrm{ell}}(z) = \Theta_p(q^{\pm 1}z)\Theta_p(t^{\mp 1}z)
    \Theta_p(q^{\mp 1} t^{\pm 1} z),
\end{equation}
The comultiplication $\Delta$ is exactly the same as in the
trigonometric case, given by Eqs.~\eqref{eq:23}. As with the matrix
model in the previous section, the essential difference with the
trigonometric case appears when one tries to build Fock representation
of elliptic DIM: one set of bosons turns out not to be enough. We need
at least \emph{two} sets of Heisenberg generators $\hat{a}_n$ and
$\hat{b}_n$ to reproduce the commutation relations of the elliptic
algebra. Concretely, we have for the level one representation:
\begin{gather}
  \rho_u (x^{+}(z)) = u \eta(z) = u\ : \exp \left( - \sum_{n \neq 0}
    \frac{(1-t^n) z^{-n}}{n (1 - q'^{|n|})} \hat{a}_n \right)
  \exp \left( - \sum_{n \neq 0}
    \frac{(1-t^{-n}) q'^{|n|} z^n}{n (1 - q'^{|n|})} \hat{b}_n \right): \notag\\
  \rho_u (x^{-}(z)) = u^{-1} \xi(z) = u^{-1} : \exp \left( \sum_{n
      \neq 0} \frac{(1-t^n) p^{-|n|/2} z^{-n}}{n (1 - q'^{|n|})}
    \hat{a}_n \right) \exp \left( \sum_{n \neq 0}
    \frac{(1-t^{-n}) p^{|n|/2} q'^{|n|} z^n}{n (1 - q'^{|n|})} \hat{b}_n \right):\notag\\
  \rho_u (\psi^{+}(z)) = \varphi^{+}(z) = \exp \left( \sum_{n > 0}
    \frac{(1-t^n) (p^{-n/2} - p^{n/2}) p^{-n/4} }{n (1 - q'^n)} \left(
      z^{-n} \hat{a}_n - p^{\frac{n}{2}} q'^n z^n
      \hat{b}_n\right)
  \right)   \label{eq:50}\\
  \rho_u (\psi^{-}(z)) = \varphi^{-}(z)= \exp \left( -\sum_{n > 0}
    \frac{(1-t^{-n}) (p^{-n/2} - p^{n/2}) p^{-n/4} }{n (1 - q'^n)}
    \left( z^n \hat{a}_{-n} - p^{\frac{n}{2}} q'^n z^{-n}
      \hat{b}_{-n} \right)
  \right)  \notag\\
  \rho_u(\gamma) = \left( t/q \right)^{1/2},\notag
\end{gather}
where $p = \frac{q}{t}$ and the bosons $\hat{a}_n$ and
$\hat{b}_n$ satisfy the following commutation relations:
\begin{gather}
  [\hat{a}_m , \hat{a}_n] = m \frac{(1 - q'^{|m|})(1 -
    q^{|m|})}{1 - t^{|m|}} \delta_{m+n,0},\notag\\
  [\hat{b}_m , \hat{b}_n] = m \frac{(1 - q'^{|m|})(1 -
    q^{|m|})}{(p q')^{|m|} (1 - t^{|m|})} \delta_{m+n,0},\label{eq:8}\\
  [\hat{a}_m , \hat{b}_n] = 0.\notag
\end{gather}
Again, the fields $\hat{a}_n$, $\hat{b}_n$ are related to
$\hat{\alpha}_n$ and $\hat{\beta}_n$ by a simple redefinition.

The \emph{dressed} current $t(z) = A(z) x^{+}(z) B(z)$, corresponding
to the stress energy tensor is given by exactly the same
expression~\eqref{eq:26}, as in the ordinary DIM case. Moreover, the
dressing operators $A(z)$ and $B(z)$ are constructed from the
$\psi^{\pm}$ generators of the elliptic DIM algebra using the same
formulas~\eqref{eq:27} as give above. In the level two representation
$\rho_{u_1, u_2}^{(2)}$ the element $t(z)$ produces the elliptic
Virasoro stress-energy tensor~\eqref{eq:12}.

Let us also mention that the \emph{undressed} elliptic DIM charge
$\oint x^{+}(z) dz/z$ also leads to several very interesting
objects. In the level one representation it gives elliptic Ruijsenaars
Hamiltonian, while in the second level representation it is the
difference version of the intermediate long-wave (ILW)
Hamiltonian~\cite{ILW}, which itself is a generalization of the
Benjamin-Ono system.

\subsection{Ward identities and $qq$-characters}
\label{sec:ward-identities-qq}

One can derive Ward identities in the same algebraic fashion as for
the trigonometric case. The OPE of the stress-energy
tensor~\eqref{eq:12} with the screening current~\eqref{eq:9} is given
by:
\begin{equation}
  \label{eq:51}
  \mathcal{T} (z) S(x) =  \frac{\Theta_{q'}\left(\frac{t
        x}{z}\right)}{\Theta_{q'}
\left(\frac{x}{z}\right)}
        : e^{\hat{\Phi}(z)}
        e^{-\hat{\Phi}(z/t)}S(x): + t \frac{\Theta_{q'}\left(\frac{q
        x}{tz}\right)}{\Theta_{q'}
\left(\frac{qx}{z}\right)} :e^{-\hat{\Phi}(tz/q)} e^{\hat{\Phi}(z/q)} S(x):
\end{equation}
which is non-singular up to total $q$-difference due to the same
cancellation, as in Eq.~\eqref{eq:7}, and we use the same time insertion operator (\ref{G}), but this time depending on two sets of times, $p_k$ and $\bar p_k$ related to sets of Heisenberg operators $\alpha_{-n}$ and $\beta_n$:
\be
{\cal G}(p)=\exp\Big(\sum_{k>0}p_{k}\alpha_{-k}+\sum_{k>0}\bar p_{k}\beta_{k}\Big)
\ee
Thus the insertion of
$\mathcal{T}(x)$ into the correlator corresponds to the insertion of
the following expression under the matrix model integral:
\be
  \label{eq:52}
\mathcal{T}_{p_k}(z)Z_{A_1}^{ell}(p) = \langle \mathcal{T}(z) \rangle =  \oint d^Nx \,
  \Delta^{(q,q',t)}_{\mathrm{ell}}(x) \left(\prod_{a=1}^M\prod_{i=1}^N V_a(z_a, x_i)U(x_i,p,\bar p)\right)\times\nn\\
  \times
  \left[ \prod_{j=1}^N \frac{\Theta_{q'}\left(\frac{t
          x_j}{z}\right)}{\Theta_{q'}\left(\frac{x_j}{z}\right)} +
    \hat{P}(z|\{z_a\}, \{ v_a\})
    \prod_{j=1}^N \frac{\Theta_{q'}\left(\frac{q
          x_j}{tz}\right)}{\Theta_{q'}\left( \frac{q x_j}{z}\right)} \right] = \hat{Q}_N(z) ,
\ee
where $\hat{P}(z|\{z_a\}, \{ v_a\})$ and $\hat{Q}_N(z)$ are products
of theta functions of the form $\prod_a \Theta_{q'}(z/\lambda_a)$, and the potential now has the form
\be
U(x,p,\bar p)=\exp \left[- \sum_{n>0} \frac{x^n}{n} \frac{1 - t^n}{(1
      - q^n)(1 - q'^{n})} \left( 1 + \left( \frac{q}{t} \right)^n \right)
    p_n - \sum_{n >0} \frac{x^{n}}{n} \frac{1 - t^{-n}}{(1
      - q^{-n})(1 - q'^{n})} \left( 1 + \left( \frac{t}{q} \right)^{n} \right)\bar p_n \right]
\ee
while the difference realization of the operator $\mathcal{T}_{p_k}(z)$ is given by the substitution
\be
\hat\alpha_{-n}=p_n,\ \ \ \ \hat\alpha_n=\frac{n (1 - q'^{n}) }{1 +
      \left( \frac{q}{t} \right)^{n}} \frac{1 - q^{n}}{1 -
      t^{n}}{\partial\over\partial p_n},\ \ \ \ \ \ \hat\beta_n=\bar p_n,\ \ \ \ \ \ \hat\beta_{-n}=-\frac{n q'^{n} (1 -
          q'^{n}) }{1 + \left( \frac{q}{t} \right)^{n}} \frac{1 -
          q^{n}}{1 - t^{n}}{\partial\over\partial \bar p_n}
\ee
This gives the elliptic
$qq$-character corresponding to the $6d$ gauge theory corresponding to
the $A_1$ quiver, i.e.\ the gauge group should consist of single
$SU(n)$ factor possibly with some fundamental matter hypermultiplets.

As we have seen in the trigonometric case, there is another very
explicit way to derive the Ward identities: to consider the vanishing
integral of a cleverly chosen total difference. In the elliptic case
this method work as well, provided the total difference is
\begin{multline}
  \label{eq:53}
  0 = \oint d^Nx \sum_{i=1}^N \frac{1}{x_i} (1 - q^{x_i \partial_i})
  \left[\sum_{k\in \mathbb{Z}} \frac{x_i t^{kN}}{z - q'^k x_i}
    \prod_{j \neq i} \frac{\Theta_{q'} \left( t\frac{x_j}{x_i}
      \right)}{\Theta_{q'} \left( \frac{x_j}{x_i} \right)}
    \Delta^{(q,q',t)}_{\mathrm{ell}}(x)
   \right] \sim \\
  \sim \oint d^Nx \, \Delta^{(q,q',t)}_{\mathrm{ell}}(x)\left[ \prod_{j=1}^N
    \frac{\Theta_{q'}\left(\frac{t
          x_j}{z}\right)}{\Theta_{q'}\left(\frac{x_j}{z}\right)} +
    t^{2N-1}q\prod_{j=1}^N
    \frac{\Theta_{q'}\left(\frac{q x_j}{tz}\right)}{\Theta_{q'}\left(
        \frac{q x_j}{z}\right)} - \hat{Q}_N (z)\right].
\end{multline}
The resulting equation is, of course the same as Eq.~\eqref{eq:52}.
The meaning of the identity~\eqref{eq:53} in the elliptic matrix model
is the same as in the $(q,t)$-matrix model: it provides the recurrence
relations for the correlators of arbitrary symmetric functions of
$x_i$. It would be interesting to obtain the factorization formulas
for the averages in this model similar to those for the averages of
(generalized) Macdonald polynomials in the $(q,t)$-model. Let us also
mention that in the Nekrasov-Shatashvili limit Eq.~\eqref{eq:53}
reduces to the quantum spectral curve of the XYZ spin chain, to the
Seiberg-Witten integrable system corresponding to the $6d$ gauge
theory.

This concludes our brief tour into the realm of elliptic matrix models
and elliptic DIM algebras. The most important lesson to learn here is
that the DIM description indeed seems to be universal: the elliptic
case is almost literally the same as the trigonometric one.

\section{Conclusions and further directions}
\label{sec:concl-furth-direct}

We have worked out the connection between a large class of network
matrix models associated with toric diagrams and the DIM algebra. The
algebra provides a unified description of the symmetry behind all such
matrix models giving rise to $qq$-characters, generalized polynomials
and Ward identities.

Application of the algebraic description to a matrix model such as
(\ref{netmod}) requires:

\begin{itemize}
\item[(i)] identification of a particular free field representation of
the appropriate DIM associated with the given model,
\item[(ii)] building explicit expressions for the screening operators expressed as integrals of screening currents
$S(x)$,
\item[(iii)] constructing the symmetry generators
(generalized stress-tensors) $\mathcal{T}(z)$ for which the screening operators are the centralizers,
\item[(iv)] representing the correlators of screening currents as Vandermonde
measures and stress-tensor insertions as $qq$-characters which can be
converted into the action of differential/difference operators. This
step relies on the $\mathcal{T}$-$S$ OPE, which should be
nonsingular up to a total difference, and the $S$-$S$ OPE,
which should give the desired version of the Vandermonde
determinant.
\end{itemize}
Schematically, one should have
\be
  \mathcal{T}(z) S(x) = \mathrm{Regular} (z, x) + (1 - q^{x\partial_x})
  \,  \mathrm{Singular} ( z/x )\,\label{eq:55}\nn \\
  S(x_1) S(x_2) = f(x_1/x_2) :S(x_1) S(x_2): \label{eq:56}
\ee
and the function $f(x)$ defines the Vandermonde factor through
$\Delta(x) = \prod_{i \neq j} f(x_i/x_j)$. For the concrete examples
of OPEs like~\eqref{eq:56} see Eqs.~\eqref{eq:21},
\eqref{eq:7}.

It is still unclear how to separate the contributions of screening
currents and vertex operators in the network matrix model formalism
since both objects are packed into a single intertwiner/topological
vertex. Probably, the technical answer to this question should depend
on the ``star-chain'' duality for conformal blocks.

This procedure is supposed to associate a D-module structure with each
particular network matrix model or, what is the same, with
representation of DIM. A non-trivial feature of actual construction,
already seen in (\ref{Tdefo}) and (\ref{eq:19}) is that the stress
tensors are actually build from roots of algebra, while the screening
operators from Cartan generators of DIM, which is somewhat against a
naive intuition coming from their realization as powers $\p\phi$ and
$\oint e^{\pm\phi}$ in the simplest free field conformal theories.
General understanding of this phenomena includes relation between the
Sugawara construction and the DIM comultiplication and between the
screening charges and the action of the Weyl group.  Remarkably, the
Weyl group of elliptic DIM should be the elliptic DAHA, of which the
elliptic Macdonald functions explicitly provided by formulas like
(\ref{sfun}) in elliptic matrix model (\ref{eq:3}), are
eigenfunctions.

An interesting question here is interpretation of the BPZ equations
\cite{CFT} for such insertions as the Baxter equations for symmetric
functions of Macdonald family, especially in elliptic case, where
there exist alternative approaches \cite{ILW}.

\begin{figure}[h]
  \centering
  \includegraphics[width=0.39\textwidth]{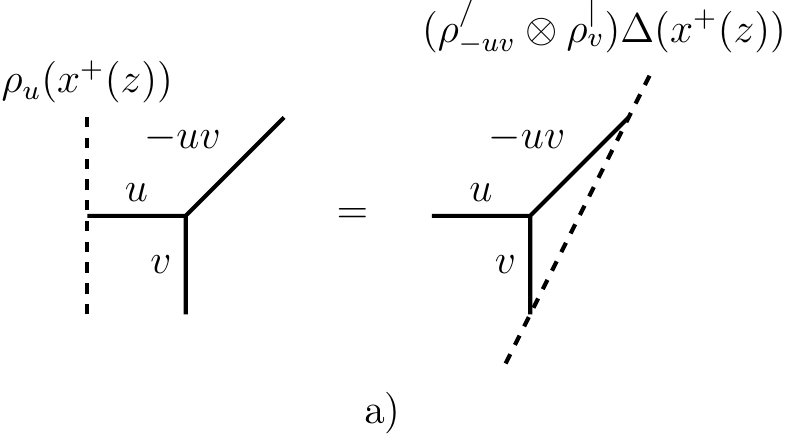}   \includegraphics[width=0.6\textwidth]{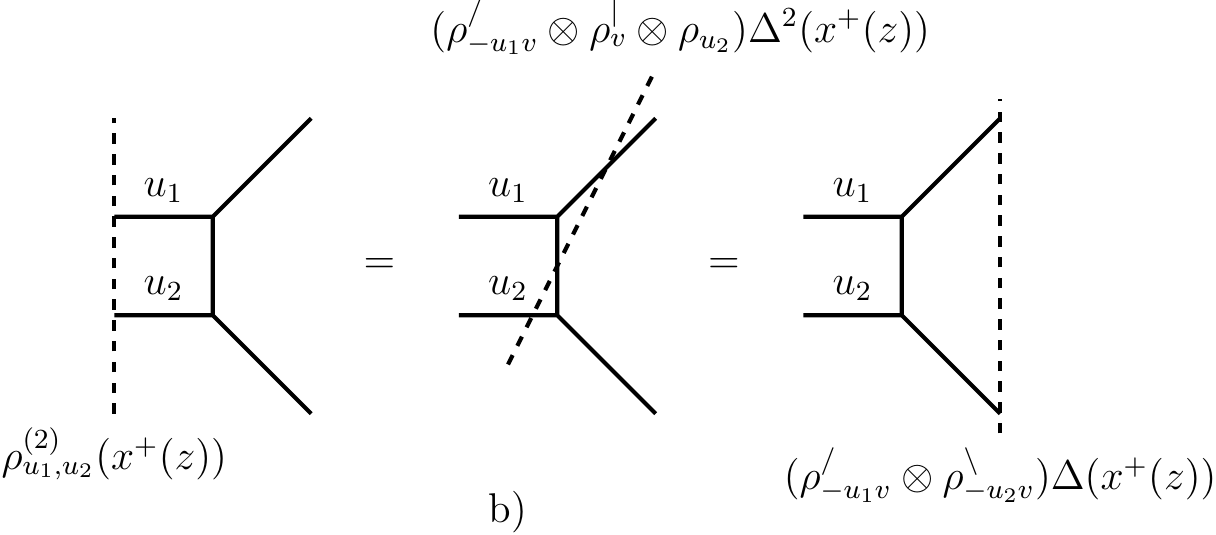}
  \caption{\footnotesize{Topological vertex as the intertwiner of DIM
    representations. a) The action of the generator $x^{+}(z)$ on the
    level one Fock representation $\rho_u$ sitting on the horizontal
    leg of the topological vertex (denoted by the dashed line) is the
    same as its action on the product of two representations --- the
    ``vertical'' $\rho_v^{|}$ and ``diagonal'' $\rho_{-uv}^{/}$. b)
    Appropriate contraction of two intertwiners is also an
    intertwiner. This gives the vertex operator of the corresponding
    conformal field theory with deformed Virasoro symmetry,
    corresponding to a single vertical brane in Fig.~\ref{fig:1}.}}
  \label{fig:2}
\end{figure}

At the level of network matrix model~\eqref{top-network} DIM symmetry
generators act on any section which cuts $M$ edges to separate the
diagram into disconnected parts (see Fig.~\ref{fig:2}). They act as
$(M-1)$-th coproduct of the original DIM generators.  As was shown
in~\cite{AF}, topological vertices are intertwiners of DIM
representations, i.e.\ the action on one of the legs is equal to the
action on two others --- this allows to pull the generator through the
vertex (Fig.~\ref{fig:2}~a)). Moreover, contraction of the legs is
consistent with this procedure (Fig.~\ref{fig:2}~b)). In the result
DIM generators can be pulled from the original section to the right of
the diagram, where negative modes annihilate the Fock vacuum, or to
the left, where the positive modes act trivially. This provides the
constraints on the matrix model averages (or, equivalently, gives the
$qq$-characters), which are the $5d$ generalization of the constraints
obtained in~\cite{Mat}. Fig.~\ref{fig:2} actually gives the uplift
of the setup considered in~\cite{Mat} to the level of the topological
string (or network matrix model). Further developments of this
approach and its applications to compactified toric diagrams will be
reported elsewhere.

\bigskip

Network matrix model is naturally built from the Seiberg-Witten
integrable system --- which is a spin chain in the simplest cases \cite{SCh,spedu}. The
network is the tropicalization of its spectral curve, and the vertical
and horizontal branes encode the rank and number of chain sites
respectively. The structure of intertwiners/$R$-matrices forming a
network can be understood as a lift of ordinary trigonometric
$R$-matrices similar to the tetrahedron
equation~\cite{Bazhanov:2005as}. This will give the connection between
the algebraic and integrable parts of the story~\cite{F}.

\bigskip

After the basic structure of network matrix model constraints is
understood, we face a multitude of different paths, each one worth
following. First of all, since DIM algebras involves double
affinization of any Lie algebra (we have only considered
$\mathfrak{gl}_1$ case) it can be applied to the affine algebra
$\widehat{\mathfrak{gl}}_1$. This should provide a \emph{triply}
affine algebra
$U_{q,t,\widetilde{t}}(\widehat{\widehat{\widehat{\mathfrak{gl}}}}_1)$
with \emph{three} parameters. In this notation it seems appropriate to
name this crucially important structure the \emph{Pagoda
  Algebra}. This algebra should have remarkable properties, one of
which is the presence of an $SL(3,\mathbb{Z})$ automorphism
group~\cite{Lockhart:2012vp}, corresponding to the automorphisms of
the compactification torus $\mathbb{T}^3$.

In the second part of this paper we have considered elliptic DIM
algebra, corresponding to $6d$ gauge theory with matter content given
by a linear quiver. The \emph{elliptization} of the \emph{triply}
affine Pagoda algebra should, therefore, describe the $6d$ gauge
theory with adjoint matter, the most mysterious of all Seiberg-Witten
systems, corresponding to double-elliptic integrable systems and
affine elliptic Selberg integrals. However, even without extra
deformations, already the case of elliptic DIM poses interesting
questions.

\bigskip

To summarize, the main idea of this paper is that DIM provides a
\emph{functor,} which lifts the \emph{picture} --- a network --- to
\emph{formulas} made out of Nekrasov functions, 3d partitions or
topological vertices. In other words, the input is a tropical spectral
curve (associated with the underlying Seiberg-Witten integrable
system) and the output is the partition function of the associated
topological string theory, which is provided by one and the same
universal procedure. At the algebraic level the input should be the
algebra $\widehat{\mathfrak{gl}}_1$ which, treated as
$\mathfrak{gl}_{\infty}$ or $W_{1+\infty}$, incorporates various
$\mathfrak{gl}_n$'s, and the output is described by the \emph{Pagoda
  algebra,} which still needs to be fully investigated.

\section*{Acknowledgements}

We are grateful to Prof.~H.~Kanno for remarkable hospitality at Nagoya
University at the last stage of this project.  We are deeply indebted
to H.~Awata, H.~Kanno, Y.~Ohkubo and V.~Pestun for lecturing us on
various aspects of the DIM symmetry and its applications and to
T.~Matsumoto and Yu.~Matsuo for encouraging comments.

Our work is partly supported by grants 15-31-20832-Mol-a-ved (A.Mor.),
15-31-20484-Mol-a-ved (Y.Z.), by RFBR grants 16-01-00291 (A.Mir.)  and
16-02-01021 (A.Mor.\ and Y.Z.), by joint grants 15-51-50034-YaF,
15-51-52031-NSC-a, 16-51-53034-GFEN, by the Brazilian National Counsel
of Scientific and Technological Development (A.Mor.).

\end{document}